\documentstyle[prl,aps,tighten,multicol]{revtex}

\begin{document}

%\draft

\title{ Enhancement of thermal entanglement in two-qubit XY models}

\author{Celia Anteneodo$^1$\thanks{\rm e-mail: celia@cbpf.br}
and Andr\'e M. C. Souza$^{1,2}$\thanks{\rm e-mail: amcsouza@.ufs.br} }
\address{$^1$ Centro Brasileiro de Pesquisas F\'{\i}sicas, \\
R. Dr. Xavier Sigaud 150, 22290-180 RJ, Rio de Janeiro, Brazil \\
         $^2$   Departamento de F\'{\i}sica, Universidade Federal de Sergipe, \\
                49100-000 SE, S\~ao Crist\'ov\~ao,  Brazil}

%\date{\today}

\maketitle

\begin{abstract}
We analyze conditions leading to enhancement of thermal entanglement in 
two-qubit XY models. 
The effect of including cross-product terms, besides the standard XY exchange 
interactions,  in the presence of an external magnetic field, is investigated. 
We show that entanglement can be yield at elevated temperatures by 
tuning the orientation of the external magnetic field. 
The details of the intrinsic exchange 
interactions determine the optimal orientation. 
 \\[3mm]
\end{abstract}

\begin{multicols}{2}

\narrowtext

%%%%%%%%%%%%%%%%%%%%%%%%%%%%%%%%%%%%%%%%%%%%%%%%%%%%%%%%%%%%%%%%%%%%%%%%
%
\section{Introduction}
Entanglement implies strong correlations between
spatially separated quantum systems that can not occur classically. 
By virtue of its nonclassical nature, 
entanglement has started to be seen in recent years as 
a phenomenon at the core of the future technology, namely, as a
potentially useful resource for quantum information processing. 
Nevertheless, beyond this exciting possibility, quantum entanglement  
deserves further investigation as it is 
not a rare phenomenon but it is generic for the states of interacting
many particle systems and, therefore, ubiquitous.

In the preparation of entangled states, the external 
control performed through changes in temperature, 
electromagnetic fields, etc., should not introduce significant levels of 
decoherence and, at the same time, the required values of the external 
control parameters must fall within realistic working ranges. 
Due to the usually large decoherence times, the spin is a property 
that has been exploited in many implementations requiring entangled states 
\cite{platzman}. 
As a consequence, most of the solid state proposals lie upon two-particle 
exchange interactions of the Heisenberg type\cite{qp016}, so that, 
the intrinsic two-body Hamiltonian operator is of the form
\begin{equation}
H'_o \;=\;  \frac{1}{2}\sum_{i=x,y,z}J^i \;\sigma_i^{(1)} \otimes  \sigma_i^{(2)}
\end{equation}
where $ \sigma_i^{(n)}$  are the Pauli  matrices of qubit $n$ and 
$J^i$ the coupling coefficients ($i=x,y,z$).
For instance, one finds the isotropic Heisenberg model 
($J^x=J^y=J^z$) in spin-coupled quantum dots\cite{s_qd} 
and donor-atom nuclear spins\cite{nes}. 
Its XY version ($J^x,J^y\neq J^z=0$) 
is found in quantum dots in  cavities\cite{qdc}, excitons in coupled 
quantum dots\cite{e_qd}, 
atoms in cavities\cite{ac} and quantum-Hall systems\cite{hall}. 
Ising-like systems ($J^x\neq J^y=J^z=0$) have also been considered\cite{ising} .

The entanglement of two-qubit systems for the XY model, 
in the presence of a  constant external magnetic field
$B_z\hat{z}$, has been investigated recently\cite{XY,XYB}. 
It was shown that the thermal entanglement of anisotropic samples 
can be manipulated through changes in $B_z$. 
It is our purpose here to analyze the effect that the inclusion of 
cross-product terms of the form $\sigma_i^{(1)}\otimes \sigma_j^{(2)}$,  
with $i\neq j$, yields upon thermal entanglement.
These terms model weak ferromagnetism and originate from 
the spin-orbit coupling\cite{qp016,crossterms}. 
Also in the scenario of fermion quantum circuits, 
the anisotropy in the exchange interaction 
may be accompanied by cross-product terms\cite{terhaldiV}. 
For external control, we will introduce a 
constant external magnetic field $B=(B_x,B_y,B_z)$. 

The degree of entanglement will be gauged through  
the {\it concurrence}\cite{concu}, a measure that, 
for two-qubit systems, gives the exact frontier between 
separable and entangled domains.  
Briefly, the concurrence $C$ is calculated as

\begin{equation}
\label{concurrence}
C\;=\;max \bigl\{ 2 max \{ \lambda_i \}-\sum_{i=1}^4 \lambda_i,\;0 \bigr\}  \;,
\end{equation}
with $\{\lambda_i\}$ the square roots of the eigenvalues  of the matrix  
$ R=\rho S \rho^* S $, where $ \rho$ is the density matrix, 
$S= \sigma^{(1)}_y \otimes \sigma^{(2)}_y$ 
and ``$*$'' stands for complex conjugate. 
A nonzero concurrence means that the two qubits are entangled, in particular  
unitary concurrence corresponds to maximally entangled states. 

The paper is organized as follows. In Section II we describe the system. 
The results are presented and discussed in Section III. Finally, Section IV 
contains the concluding remarks. 

%%%%%%%%%%%%%%%%%%%%%%%%%%%%%%%%%%%%%%%%%%%%%%%%%%%%%%%%%%%%%%%%%%%%%%%%
%
\section{The system}

In order to study the effect of extra cross-product terms in the XY model, 
let us consider the following internal Hamiltonian  

\begin{equation} \label{ham0}
H_o  \; =\; \frac{1}{2} \sum_{i,j=x,y} a^{ij}\; \sigma_i^{(1)} \otimes \sigma_j^{(2)},
\end{equation}
with  $a^{xx}=(1+\gamma)J$, $a^{yy}=(1-\gamma)J$, where $J\in \Re$ is the coupling 
constant and  $\gamma \in [-1,1]$ is an anisotropic parameter. 
Also, let us define the cross couplings as 
$a^{xy}=(1+\gamma')K$ and $a^{yx}=(1-\gamma')K$ with $K \in \Re$ the 
cross coupling constant and $\gamma' \in [-1,1]$. 
Then, $H_o$, written in terms of the raising and 
lowering operators $\sigma^{(n)}_\pm=(\sigma^{(n)}_x \pm i\sigma^{(n)}_y)/2$, reads
\begin{eqnarray} \nonumber
H_o &=&            (\gamma J-i K)\;\sigma_{+}^{(1)}\otimes\sigma_{+}^{(2)} 
                     \;+\;(J+i\gamma' K)  \;\sigma_{+}^{(1)}\otimes\sigma_{-}^{(2)} 
              \\          
                &+&(J-i\gamma' K)  \;\sigma_{-}^{(1)}\otimes\sigma_{+}^{(2)}
                 \;+\;(\gamma J+i K) \;\sigma_{-}^{(1)}\otimes\sigma_{-}^{(2)} .
\label{ham00}
\end{eqnarray}
It is clear now that the inclusion  of cross-product terms 
is equivalent to setting complex coupling constants.

In the presence of the magnetic field $B$,  
the complete dimensionless Hamiltonian becomes
\begin{equation} \label{ham}
H  \; =\; H_o \;+\;\frac{1}{2} \sum_{i=x,y,z} B_i( \sigma_i^{(1)} \oplus \sigma_i^{(2)} ). 
\end{equation}

We consider states of equilibrium at temperature $T$. 
Thus, the density  matrix is given by 
$\rho=\exp(-H/T)/Z$ with $Z=Tr(\exp(-H/T))$ where 
we have already set the Boltzmann constant $k_B=1$. 
The entanglement at these states is called {\em thermal entanglement}\cite{thentang}.

%%%%%%%%%%%%%%%%%%%%%%%%%%%%%%%%%%%%%%%%%%%%%%%%%%%%%%%%%%%%%%%%%%%%%%%%
%
\section{Results}

Let us analyze first the case $B_\perp=B_z\neq 0, B_\parallel=0$, where 
$B_\parallel$ is the component of the field in the $xy$-plane. 
The eigenstates of the density matrix $\rho$ are
\begin{equation} \label{psi}
\mid \psi_\mu^\pm\rangle \;=\; \bigl[  (J+i\gamma' K) \mid\uparrow\downarrow \rangle \,\pm\, 
\mu \mid\downarrow\uparrow\rangle \bigr]/N_\mu
\end{equation}
(with eigenvalues $\pm \mu$, respectively, being $\mu=\sqrt{J^2+(\gamma'K)^2}\;\;$) and 
\begin{equation} \label{phi}
\mid \phi_\nu^\pm\rangle \;=\; 
\bigl[ (\gamma  J-i K) \mid\uparrow\uparrow\rangle \,+\, 
(\pm \lambda-B_\perp ) \mid\downarrow\downarrow\rangle \bigr]/N_\lambda
\end{equation}
(with eigenvalues $\pm\lambda$, being  $\lambda=\sqrt{\nu^2+B_\perp^2}$, 
 $\nu^2= (\gamma J)^2+K^2\;\;$), where $N_\mu, N_\lambda$ are 
normalization constants. 

The eigenvalues of matrix $R$ are given by
\begin{eqnarray} \nonumber
\lambda^2_{1,2}&=& \exp(\pm 2 \mu /T )/Z^2 \\ \label{lR}
\lambda^2_{3,4}&=& (1\,+\, 2x^2 \pm 2x \sqrt{1+x^2} )/Z^2   
\end{eqnarray}
with $x= (\nu/\lambda)\, \sinh(\lambda/T)$   
and $Z=2\cosh(\nu/T)+2\cosh(\lambda/T)$.
Then, the concurrence $C$ is straightforwardly computed through Eq. (\ref{concurrence}). 
At $T\to\infty$ or $B_\perp\to\infty$, the entanglement vanishes ($C\to 0$).
At $T=0$, the concurrence is
\begin{equation} \label{CT0}
C(T=0) \;=\; \left\{ \matrix{ 
    1                 && \mu>\lambda,  \cr
(1-\nu/\lambda)/2   &\mbox{for} & \mu=\lambda, \cr
   \nu/\lambda        && \mu<\lambda . } \right.
\end{equation}
Hence, for $T=0$, there is a jump in the concurrence at $\mu=\lambda$ 
similarly as already observed in the literature for the case $K=0, 
\gamma\neq 0$\cite{XYB}. 
For $\mu>\lambda$, the ground state is $\mid \psi_\mu^-\rangle$, 
which is fully entangled. 
For $\mu<\lambda$, the ground state, $\mid \phi_\nu^-\rangle$,  
is partially entangled unless $B_\perp=0$ in which case it becomes fully 
entangled. 
But, at $\mu=\lambda$, the ground  state disentangles 
partially(totally) for $B_\perp \neq 0$(=0).

The results are not dependent on the sign of the coupling parameters, 
therefore, in particular, they are valid both for the antiferromagnetic 
and ferromagnetic cases. Also they  are independent of the sign of $B_\perp$. 

The picture in the absence of magnetic field ($|B|=0$) is recovered 
by setting $B_\perp=0$ in expressions (\ref{psi})-(\ref{CT0}). 
In Fig. 1 we show $T^*$, the temperature at which the 
concurrence vanishes, as a function of parameters $\gamma'$, for different 
sets of $(J,K)$, in the absence of any external magnetic field. 
The concurrence is null above $T^*$. 
If $\gamma'^2>1-(J/K)^2$, then, the threshold temperature $T^*$ grows with  
increasing $\gamma'$. 
For $|B|=0$, by interchanging the values of parameters $(K,\gamma')$ with those 
of $(J,\gamma)$, one obtains the same results. 
Taking this into account, notice that, even for the Ising case ($|\gamma|=1$), 
there can be entanglement at finite temperature in the absence of magnetic 
field ($|B|=0$), as soon as $K$ is large enough. 

Fig. 2 exhibits the effect of the external field $B_\perp$ on 
an isotropic sample ($\gamma=0$). 
By increasing $|B_\perp|$, it is possible to increase the threshold 
temperature $T^*$ provided $K\neq0$. 
In reference \cite{XYB} it was shown that anisotropic exchange interactions 
are required for allowing control of $T^*$ by varying the intensity of $B_\perp$. 
However, the presence of cross-terms enables control of the  threshold temperature 
by varying the intensity of the magnetic field in the $z$-direction, 
{\it even in the absence of anisotropy}.  

If $B_\parallel = 0$, the subspaces spanned by 
$\{\mid\uparrow\uparrow\rangle, \mid\downarrow\downarrow\rangle\}$ and  by
$\{ \mid\uparrow\downarrow\rangle, \mid\uparrow\downarrow\rangle\}$ 
remain uncoupled, while for $B_\parallel\neq0$, they mix. 
Numerical calculations are performed in this case. 
As expected, the results are only dependent on the total component 
of the magnetic field in the $xy$-plane, $B_\parallel$. 
Plots of the concurrence as a function of $T$ for different values of 
$B_\parallel$, being  $B_\perp=0$, are presented in Fig. 3.a. 
For $T=0$, there is a jump in the concurrence at $B_\parallel=\sqrt{2}J$ when 
$\gamma,K=0$.  In this case, one obtains
\begin{equation} \label{CT0parallel}
C(T=0) \;=\; 
\left\{ \matrix{ 
   1                        && B_\parallel<\sqrt{2}J, \cr
   2/3                       & \mbox{for} & B_\parallel=\sqrt{2}J, \cr
   \frac{1-D^2}{1+D^2}      && B_\parallel>\sqrt{2}J , } \right.
\end{equation}
with $D=(\sqrt{1+(2B_\parallel/J)^2}-1)/(2B_\parallel/J) $. 

For $B=B_\perp \hat{z}$ and $\gamma,K=0$, it was already shown in the 
literature \cite{XY} that, when $B_\perp>J$, although there is no entanglement 
($C=0$) at $T=0$ (also in agreement with Eq. (\ref{CT0})), the 
concurrence presents a maximum at finite $T$. 
In fact, at $T=0$, the ground state is the disentangled state 
$|\downarrow\downarrow\rangle$, that mixes with the maximally entangled 
ones as $T$ increases. 
A similar effect is observed for $B=B_\parallel$ and $\gamma,K=0$ (see Fig. 3.a), 
although in this case  $C(T=0) \neq 0$ consistently with Eq. (\ref{CT0parallel}).  
For $B_\parallel<\sqrt{2}J$, the concurrence $C$ decreases monotonously with $T$, 
whereas,  for $B_\parallel>\sqrt{2}J$, it presents a local maximum at finite $T$.  

A plot of $C$ vs. $|B|$ is shown in Fig. 3.b for different 
orientations, at $T\simeq 0$ and $\gamma,K=0$. 
For all orientations there is a jump at $T=0$. It occurs when  

\begin{equation} \label{cond}
2B^2_\perp + B^2_\parallel =2 J^2.
\end{equation}
Differently from the case $B=B_\perp \hat{z}$, if $B_\parallel\neq 0$, then, 
the concurrence falls to a non null value above the singularity.  
Thereafter, $C$ tends smoothly to zero with increasing intensity of the 
magnetic field.

In Fig. 4 we exhibit $T^*$ as a function of the intensity $|B|$ for 
different orientations of the field when $\gamma=0$ and $K=0$. 
Notice that if the magnetic field is oriented in the $z$-direction, 
then  the threshold temperature does not depend on the intensity of the field. 
However, when the field has a non null component in the $xy$-plane 
($B_\parallel\neq 0$), then, $T^*$ increases for sufficiently large $B_\parallel$.  
By comparing Figs. (2) and (4) notice that $B_\parallel\neq0$ yields an
effect qualitatively similar to that due to $B_\perp$ when the system 
has cross-terms ($K \neq0$).  A similar effect is 
also observed with $B_\perp$ when the system is anisotropic ($\gamma \neq 0$)\cite{XYB}.  
But, even in the absence of anisotropy or cross-terms the external magnetic 
field can be used to control the entangled domain by 
switching on the $\parallel$-component. 

%%%%%%%%%%%%%%%%%%%%%%%%%%%%%%%%%%%%%%%%%%%%%%%%%%%%%%%%%%%%%%%%%%%%%%%%
%
\section{Conclusions}

To summarize, we have investigated the thermal entanglement of 
two-qubits XY systems with additional exchange cross-terms ($K\neq 0$) 
as well as with anisotropic couplings ($\gamma\neq0$). 
Arbitrary orientations of an external magnetic field $B$ were considered. 
We have found that, {\it even in the absence of anisotropy}, cross-terms 
enables control of the threshold temperature $T^*$, which defines the domain 
of entangled states, by varying the intensity of the magnetic field in the 
$z$-direction. 
Moreover, even in the absence of anisotropy 
or cross couplings, the degree of entanglement can be manipulated by 
suitably changing the relative orientation of the sample with respect to 
a constant magnetic field. 
As a perspective, it would be interesting to investigate the effect of 
increasing the number of spins. 

Since thermal entanglement is a natural type of entanglement for a 
system embedded in a thermal environment, 
we expect that the present results could be useful for solid state 
applications. 
In particular we expect that the understanding of  
thermal entanglement could be exploited within the context 
of quantum communication and information processing, e.g., 
along the lines of quantum heat machines \cite{qengines}.

\section*{Acknowledgements} 
A.M.C.S. is grateful to C. Tsallis for his 
kind hospitality at CBPF where this work was done. 
We acknowledge Brazilian Agencies FAPERJ and  CNPq  for financial
support.

\section*{Figure captions}
\mbox{}
\\[3mm] 
{\bf Figure 1:} Temperature $T^*$ above which the 
concurrence vanishes as a function of $\gamma'$,  
in the absence of magnetic field $|B|=0$, for $\gamma=0$ and 
different values of $(J,K)$ indicated in the figure. 
\\[3mm]
{\bf Figure 2:} Temperature $T^*$ as a function of $B_\perp$, for the parameters 
indicated in the figure. The entangled domain is the low-temperature region 
bounded by the curve and abscises axis.
\\[3mm] 
{\bf Figure 3:} Concurrence $C$ as a function of $T$ (a) for different values of 
$B_\parallel$ and $B_\perp=0$ and $C$ as a function of $|B|$ at $T\simeq 0$ (b) for 
different orientations of the magnetic field.
 $B_\perp=|B|\cos\theta$ and $B_\parallel = |B|\sin\theta$, 
where the values of angle $\theta$ are indicated in the figure. 
Here $\gamma=0$, $K=0$.
\\[3mm] 
{\bf Figure 4:} Temperature $T^*$ vs $|B|$ for the orientations of the field 
indicated in the figure. The entangled domain is the low-temperature one. 
Here $\gamma=0$, $K=0$.
%%%%%%%%%%%%%%%%%%%%%%%%%%%%%%%%%%%%%%%%%%%%%%%%%%%%%%%%%%%%%%%%%%%%%%%%

\end{multicols}

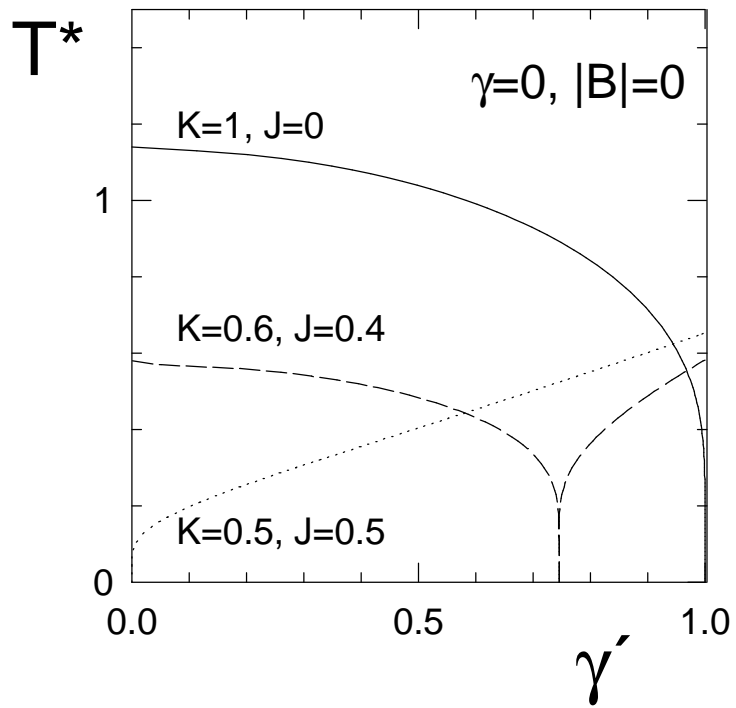
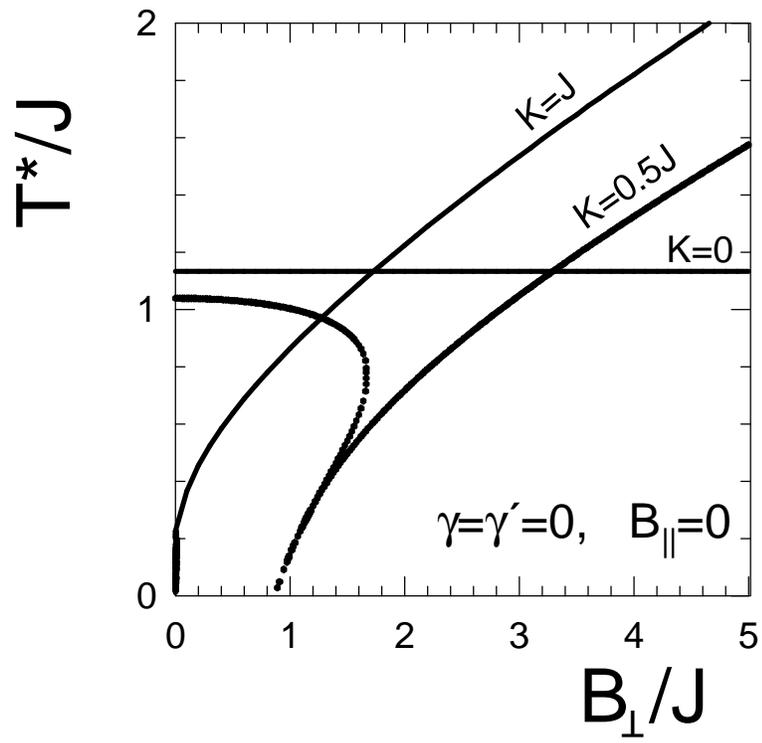
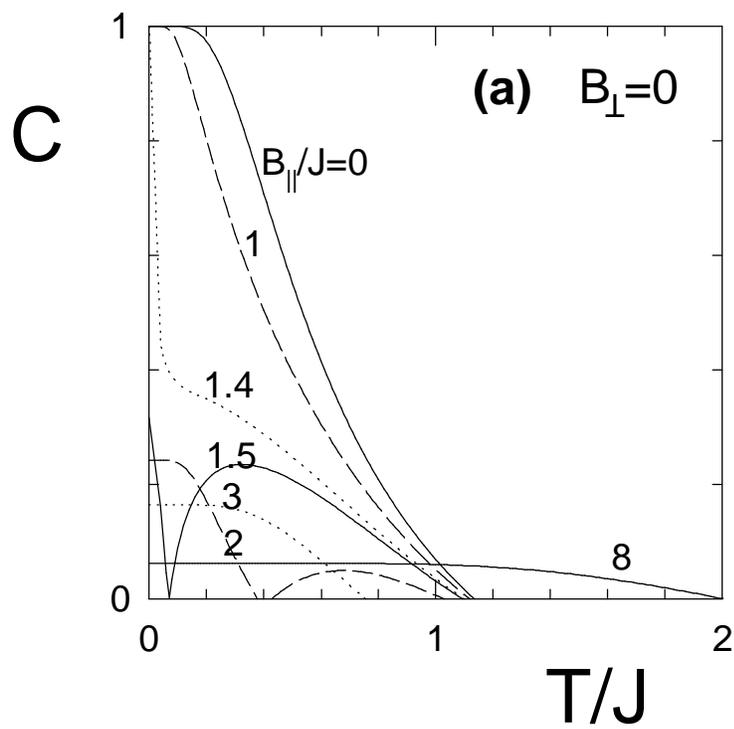
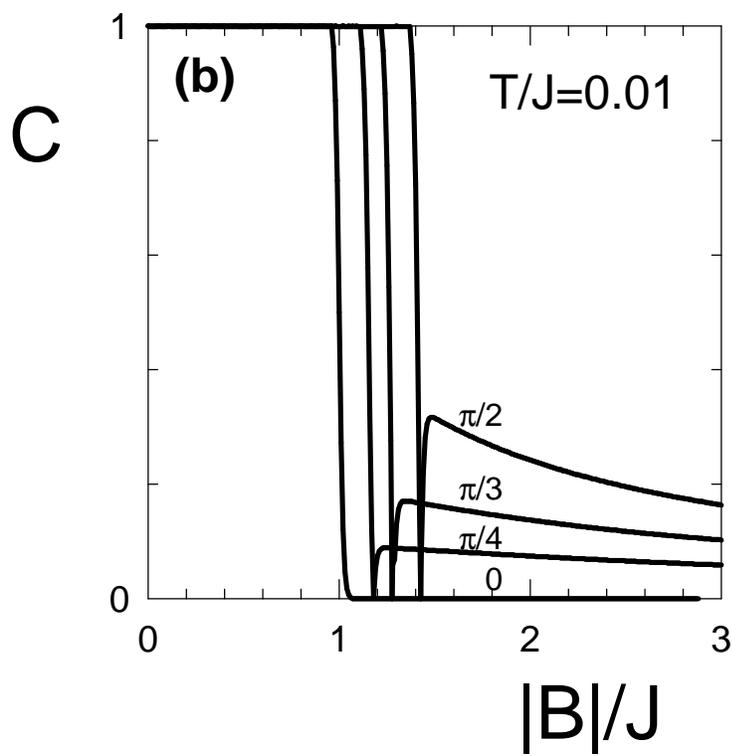
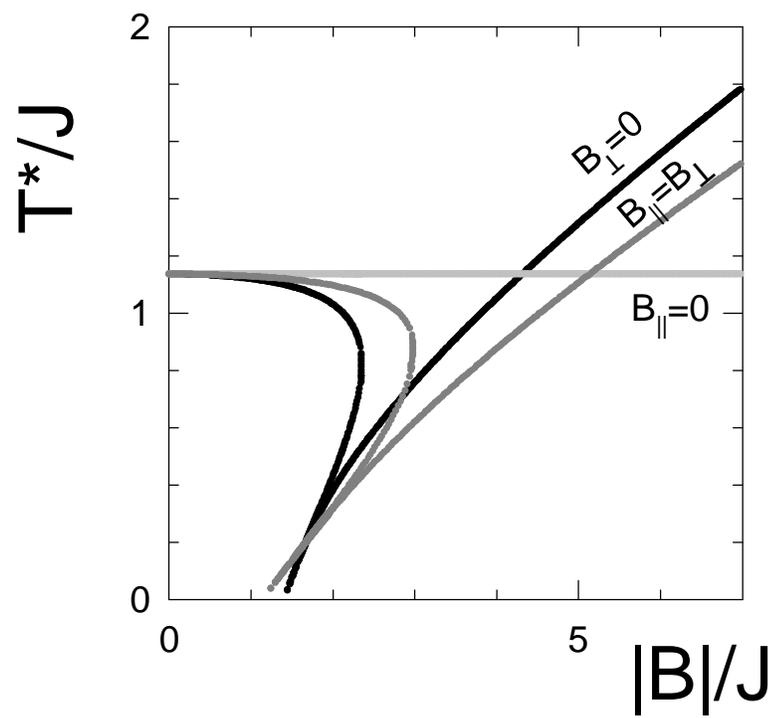
\end{document}